\documentclass[aps,prd,10pt,superscriptaddress,twocolumn,shortbibliography,showkeys]{revtex4-1}

\usepackage{graphicx}
\usepackage{xcolor}
\usepackage{amsfonts}
\usepackage{amsmath}
\usepackage{amssymb}
\usepackage{hyperref}
\usepackage{multirow}
\usepackage{enumitem}
\usepackage{subcaption}
\usepackage[sort&compress]{natbib}

\usepackage[table, xcdraw]{xcolor}
\usepackage{colortbl}
\definecolor{hred}{RGB}{255,220,220}  
\definecolor{hblue}{RGB}{220,235,255}  

\hypersetup{colorlinks=true, linkcolor=blue, urlcolor=blue, citecolor=blue}

\begin{document}

\title{Quintessential Implications of the presence of AdS in the Dark Energy sector}

\author{Purba Mukherjee} 
\email{pdf.pmukherjee@jmi.ac.in}
\affiliation{Centre for Theoretical Physics, Jamia Millia Islamia, New Delhi-110025, India}%

\author{Dharmendra Kumar}
\email{2019rpy9096@mnit.ac.in}
\affiliation{Malaviya National Institute Of Technology Jaipur-302017, Rajasthan, India}

\author{Anjan A Sen}
\email{aasen@jmi.ac.in}
\affiliation{Centre for Theoretical Physics, Jamia Millia Islamia, New Delhi-110025, India}%

% \date{\today}

\begin{abstract}
We explore the implications for an Anti-de Sitter (AdS) vacuum, equivalently a negative cosmological constant (nCC), in the dark energy (DE) sector using current cosmological observations. Our joint analysis uses DESI BAO, DESY5 supernovae, and P-ACT CMB (ACT-DR6 + Planck) measurements. We also use the KiDS weak-lensing measurement to cross-check consistency with the inferred value of $S_{8}$. Within the Chevallier--Polarski--Linder parametrization for the evolving component of the DE, the inclusion of an AdS term provides a theoretically motivated mechanism that opens up a finite non-phantom region in the CPL parameter space while remaining compatible with current observations. A negative cosmological constant also implies a finite cosmic lifetime, thereby linking observational DE phenomenology to broader questions in quantum gravity and string theory.
\end{abstract}


\maketitle

\section{Introduction} \label{sec:intro}

Over the past few decades, cosmology has made significant progress, largely due to precise observations from the Cosmic Microwave Background (CMB) \cite{Planck:2018vyg,Tristram:2023haj}, Type Ia supernovae (SN-Ia) \cite{Brout:2022vxf, DES:2024jxu, Rubin:2023ovl}, large-scale structure (LSS) and galaxy surveys \cite{eBOSS:2020yzd, DESI:2024jis, DESI:2024lzq, DESI:2024uvr}. These have improved our understanding of the universe's composition, expansion, and structure. The standard model of cosmology, $\Lambda$CDM, suggests that the universe is primarily made up of cold dark matter (CDM) and dark energy (DE), described by a positive cosmological constant \cite{Zeldovich:1968ehl}($\Lambda$) responsible for the universe's accelerated expansion \cite{Weinberg:1988cp,  Sahni:1999gb}. However, recent observations have revealed some key issues \cite{Hazra:2013dsx, Verde:2019ivm, Riess:2021jrx}, such as the Hubble tension \cite{DiValentino:2020zio} ($>5\sigma$ disagreement between local measurements of the Hubble constant $H_0$ and CMB-based estimates), and the $\sigma_8$ tension \cite{DiValentino:2020vvd} ($\approx 2.5\sigma$ mismatch between the CMB-predicted and observed clustering of matter from galaxy surveys \cite{DES:2021wwk, Heymans:2020gsg, Li:2023tui}). Furthermore, deep-space James Webb Space Telescope (JWST) observations have identified the existence of massive and bright galaxies at redshifts $z \gtrsim 7$ \cite{Labbe:2022ahb, Boylan-Kolchin:2022kae}, which is difficult to obtain with standard $\Lambda$CDM model. These discrepancies point to potential gaps in our understanding of the universe's expansion and structure formation, suggesting the need for modifications to the 6-parameter baseline $\Lambda$CDM framework \cite{DiValentino:2021izs, Perivolaropoulos:2021jda,  Abdalla:2022yfr}

The nature of dark energy (DE) remains one of the most profound open problems in cosmology \cite{Sahni:2006pa, Huterer:2017buf}. Recent baryon acoustic oscillation (BAO) data from the Dark Energy Spectroscopic Instrument (DESI) suggests that the DE equation of state (EoS) may evolve with cosmic time, with indications of early phantom-like behaviour \cite{DESI:2025zgx, DESI:2025fii}. While the $\Lambda$CDM model provides a good fit to DESI data, extending the model to CPLCDM shows deviations from $w=-1$ {at $3.1\sigma$ confidence level (CL) in combination with Planck CMB data.} Combining these observations with Pantheon-Plus (or DES-5YR) SN-Ia data, the CPLCDM model provides a better fit, excluding $\Lambda$CDM at $\sim2.8\sigma$ (or $4.2\sigma$) CL. The data suggest a preference for an early-phantom to late-non-phantom transition in the DE EoS, prompting the community to search for alternative scenarios for explaining the cosmic dark sector \cite{DESI:2024kob, Cortes:2024lgw, Chan-GyungPark:2024mlx, Wolf:2023uno, Gialamas:2024lyw, Jiang:2024xnu, Dinda:2024kjf, Wang:2024hks, Colgain:2024xqj, Bhattacharya:2024hep, Ferrari:2025egk, Chan-GyungPark:2025cri, Wolf:2024stt, Ghosh:2024kyd, Ye:2024ywg, Mukherjee:2024ryz, Tiwari:2024gzo, Mukherjee:2024pcg, RoyChoudhury:2024wri, Sakr:2025daj, Sakr:2025fay, Mukherjee:2025ytj}.

Another intriguing possibility is that the DE sector contains an Anti–de Sitter (AdS) vacuum — equivalently, a negative cosmological constant — together with an evolving component. Such scenarios are theoretically motivated by string theory and AdS/CFT correspondence, and have profound implications for cosmic dynamics, including a finite lifetime for the Universe. Such models have gained interest in recent times \cite{Cardenas:2002np, Dutta:2018vmq}. Although constructing de Sitter (dS) vacua in string theory is challenging \cite{Vafa:2005ui, Danielsson:2018ztv, Obied:2018sgi, Garg:2018reu, Cicoli:2018kdo}, AdS vacua can be more easily realized within the string landscape \cite{Maldacena:1997re}. Recent studies have shown that AdS vacua may help in addressing the cosmological tensions and discrepancies observed in the CMB, BAO, and even JWST data \cite{Akarsu:2019hmw, Visinelli:2019qqu, Calderon:2020hoc, Sen:2021wld, Ye:2020btb, Menci:2024rbq, Gomez-Valent:2023uof, Ruchika:2020avj, Jiang:2021bab, Adil:2023exv, Wang:2024hwd, Menci:2024hop}. While a nCC alone cannot drive the universe's accelerated expansion, it could coexist with a positive DE component, providing a potential framework for understanding the nature of dark energy. 

In this \textit{letter}, we test for the presence of an AdS contribution using a joint analysis of DESI BAO, DESY5 supernovae, and P-ACT CMB data, with KiDS weak lensing for cross-validation. Modeling the evolving DE with the Chevallier-Polarski-Linder (CPL) \cite{Chevallier:2000qy, Linder:2002et} form, allowing for an AdS minimum is compatible with current observational constraints and show that its inclusion naturally permits non-phantom DE evolution, avoiding field-theoretically problematic phantom behavior. These results indicate that a negative cosmological constant in the DE sector remains observationally viable, and also allow for DE evolution that remain non-phantom within the $2\sigma$ significance, suggesting that scalar field theories can effectively model DE evolution without violating the null energy condition (NEC) \cite{Ratra:1987rm, Caldwell:1997ii}. We also study the finite lifetime of our Universe in such DE models with AdS minima.

\section{Theoretical Framework} \label{sec:theory}

We outline the cosmological framework for a spatially flat Friedmann-Lema\^{i}tre-Robertson-Walker (FLRW) universe, where the Hubble parameter $H(z)$ is given by 
\begin{equation}
H^2(z) = H_0^2 \left[\Omega_m(1+z)^3 + \Omega_r(1+z)^4 + \Omega_{\rm DE}(z)\right] \, . \end{equation}
Here $z$ is the redshift, $H_0$ is the Hubble parameter, and $\Omega_i$ [``i" stands for matter ($m$), radiation ($r$) and dark energy (DE)] are the density parameters at the present epoch. For convenience, we define the reduced Hubble parameter $E(z)$ as,
\begin{equation}
\small E(z) \equiv \dfrac{H(z)}{H_0}  =  \left[\Omega_m(1+z)^3 + \Omega_r(1+z)^4 + \Omega_{\rm DE}(z)\right]^{\frac{1}{2}}
\end{equation} that characterizes the expansion rate of the Universe.
We assume the presence of nCC ($\Lambda <0$) in the DE sector, i.e. total DE density, $\rho_{\rm DE} = \rho_\phi + \Lambda$. $\rho_\phi$ is the evolving part of DE, which in principle can be modeled by a minimally coupled slow-rolling
scalar field \cite{Ratra:1987rm, Caldwell:1997ii}. $\rho_{\rm DE}$ must remain positive to give late time acceleration and possibly throughout the history of the Universe up to present time to avoid any collapsing Universe in the past. With this, the density parameter of DE is expressed as
\begin{equation}
\Omega_{\rm DE}(z) = \Omega_\phi \, f(z) + \Omega_\Lambda \, ,
\end{equation}
where $f(z)$ captures the evolving part of the DE and is defined as $f(z) = \exp\left[3 \int_0^z \frac{1+w_\phi(z')}{1+z'} \, \mathrm{d}z'\right].$ The flatness condition requires $\Omega_m + \Omega_r + \Omega_\phi + \Omega_\Lambda = 1$. At present, the combined DE sector satisfies,
\begin{equation}
\Omega_{\rm DE}(z=0) = \Omega_\Lambda + \Omega_\phi \approx 1 - \Omega_m - \Omega_r \, .    
\end{equation}
Without specifying any specific field theory model for the evolving part of the DE, we adopt two parametrizations for $w_\phi = p_\phi / \rho_\phi$ viz. (i) $w_\phi = w_0$ (constant), (ii) $w_\phi = w_0 + w_a \left(\frac{z}{1+z}\right)$ (i.e., CPL parametrization  \cite{Chevallier:2000qy, Linder:2002et}). These models correspond to the $w$CCCDM and CPLCCCDM frameworks in our study. Note that both of these models for DE include conventional {$w$CDM} and {CPLCDM} models as considered in the DESI-DR2 \cite{DESI:2025zgx} analysis for $\Omega_{\Lambda} = 0$. We aim to study whether the presence of $\Lambda$, in particular the negative $\Lambda$ (nCC) in the DE sector, brings novel signatures in the expansion history of the universe. Furthermore, we seek to explore the effect of nCC in addressing different cosmological tensions. \smallskip

\squeezetable
\begin{table}[t]
\setlength{\tabcolsep}{15pt}
\centering
\begin{minipage}{0.5\textwidth}
\centering
\resizebox{\textwidth}{!}{\renewcommand{\arraystretch}{1.3}
\begin{tabular}{lcc}
\hline\hline
\multicolumn{3}{c}{\textbf{KiDS}} \\
\hline
\textbf{Parameter} &    \textbf{$w$CCCDM}  &  \textbf{CPLCCCDM} \\
\hline
{\boldmath$\omega_{\rm b}$}        & $0.0226\pm 0.0020$                    & $0.0225\pm 0.0020$          \\
{\boldmath$\omega_{\rm cdm}$}        & $0.122^{+0.041}_{-0.053}$      & $0.130^{+0.047}_{-0.054}$   \\
{\boldmath$h$}                         & $0.712^{+0.048}_{-0.058}$          & $0.727\pm 0.051$            \\
{\boldmath$\sigma_{8,0}$}  & $0.83\pm 0.12                        $ & $0.81\pm 0.13              $\\
{\boldmath$n_s$}                       & $0.968\pm 0.076$           & $0.957\pm 0.076$            \\
{\boldmath$w_0         $}   & $-0.88\pm 0.17             $   & $-0.79^{+0.40}_{-0.25}     $\\
{\boldmath$w_a         $}           &        -            & $-1.2^{+1.1}_{-1.0}        $\\
{\boldmath$\Omega_{\rm \phi}$}    & $2.21^{+0.37}_{-0.55}$             & $2.1^{+1.1}_{-1.3}$         \\
\hline 
{\boldmath$\Omega_{\rm m}        $}   & $0.288^{+0.059}_{-0.099}   $   & $0.291^{+0.087}_{-0.10}    $\\
\multicolumn{3}{>{\columncolor{yellow!30}}c}{\strut} \\[-3.8ex]
{\boldmath$\Omega_\Lambda   $} &      $-1.50^{+0.63}_{-0.38}     $ &             $-1.4^{+1.6}_{-2.3}        $\\
\multicolumn{3}{>{\columncolor{hred}}c}{\strut} \\[-3.8ex]
{\boldmath$H_0              $}   & $71.2^{+3.7}_{-6.1}   $   & $72.7^{+4.9}_{-6.1}   $\\
{\boldmath$S_8            $} &   $0.792^{+0.050}_{-0.045}   $ & $0.774^{+0.079}_{-0.088}   $\\
\hline
{\boldmath$\chi^2_{\rm min}$}      & $81.371$ & $81.497$  \\
\hline
{\boldmath $\Delta \chi^2$ } &  $-1.565$ & $-1.771$\\

\hline
\multicolumn{3}{c}{\textbf{P-ACT+DESI}} \\
\hline
\textbf{Parameter} &    \textbf{$w$CCCDM} &     \textbf{CPLCCCDM} \\
\hline
{\boldmath$\omega_b$} & $0.02254\pm 0.00011$ &  $0.02249\pm 0.00011$ \\
{\boldmath$\omega_{\rm cdm}$}  & $0.11761\pm 0.00092$  & $0.1190\pm 0.0011$ \\
{\boldmath$100 \, \theta_{s}$}  & $1.04178\pm 0.00025$  & $1.04169\pm 0.00025$ \\
{\boldmath$\ln \left(10^{10}A_s\right)$}  & $3.061\pm 0.011$  & $3.051\pm 0.011$ \\
{\boldmath$n_s$}  & $0.9744\pm 0.0034$ & $0.9714\pm 0.0035$ \\
{\boldmath$\tau_{\rm reio}$}  & $0.0610\pm 0.0064$ & $0.0591\pm 0.0060$ \\
\multicolumn{3}{>{\columncolor{orange!20}}c}{\strut} \\[-3.8ex]
{\boldmath$w_{0}$} & $-1.013^{+0.024}_{-0.020}$  & $-0.71^{+0.16}_{-0.20}$ \\
\multicolumn{3}{>{\columncolor{orange!20}}c}{\strut} \\[-3.8ex]
{\boldmath$w_{a}$} &  -  &  $-0.84^{+0.56}_{-0.44}$ \\
{\boldmath$\Omega_{\rm \phi}$}  & $1.75^{+0.71}_{-1.1}$ & $1.40^{+0.54}_{-0.99}$ \\
\hline 
{\boldmath$\Omega_{\rm m}$}   & $0.2950\pm 0.0076$  & $0.344\pm 0.022$ \\
\multicolumn{3}{>{\columncolor{yellow!30}}c}{\strut} \\[-3.8ex]
{\boldmath$\Omega_\Lambda$}  & $-1.04^{+1.1}_{-0.71}$ & $-0.74^{+0.98}_{-0.55}$ \\
\multicolumn{3}{>{\columncolor{hblue}}c}{\strut} \\[-3.8ex]
{\boldmath$H_0$}   & $69.11\pm 0.99$  & $64.4\pm 2.0$ \\
{\boldmath$S_8$} &   $0.8121\pm 0.0076$  & $0.842\pm 0.014$ \\
\hline
{\boldmath$\chi^2_{\rm min}$}      & $826.186$ & $822.749$ \\
\hline
{\boldmath $\Delta \chi^2 $}  &  $2.576$ & $2.469$ \\

\hline
\multicolumn{3}{c}{\textbf{P-ACT+DESI+DESY5}} \\
\hline
\textbf{Parameter}  &  \textbf{$w$CCCDM}  &  \textbf{CPLCCCDM} \\
\hline
{\boldmath$\omega_b$}           & $0.02258\pm 0.00011$        & $0.02250\pm 0.00011$        \\
{\boldmath$\omega_{\rm cdm}$}     & $0.11650\pm 0.00083$         & $0.1186\pm 0.0010$          \\
{\boldmath$100 \, \theta_{s}$}      & $1.04186\pm 0.00024$          & $1.04171\pm 0.00025$        \\
{\boldmath$\ln \left(10^{10}A_s\right)$}         & $3.068\pm 0.012$                 & $3.053\pm 0.011$            \\
{\boldmath$n_s$}          & $0.9768\pm 0.0033$            & $0.9723\pm 0.0034$          \\
{\boldmath$\tau_{\rm reio}$}   & $0.0624\pm 0.0067$   & $0.0591\pm 0.0058$          \\
\multicolumn{3}{>{\columncolor{orange!20}}c}{\strut} \\[-3.8ex]
{\boldmath$w_0$}       & $-0.978^{+0.013}_{-0.017}$    & $-0.877^{+0.044}_{-0.064}$  \\
\multicolumn{3}{>{\columncolor{orange!20}}c}{\strut} \\[-3.8ex]
{\boldmath$w_a$}   &     -     & $-0.39^{+0.22}_{-0.16}$     \\
{\boldmath$\Omega_{\rm \phi}$}&   $2.1^{+1.0}_{-1.3}$                & $1.43^{+0.49}_{-0.35}$      \\
\hline 
{\boldmath$\Omega_{\rm m}$}              & $0.3094\pm 0.0052$                & $0.3173\pm 0.0060$          \\
\multicolumn{3}{>{\columncolor{yellow!30}}c}{\strut} \\[-3.8ex]
{\boldmath$\Omega_\Lambda$}  &    $-1.4^{+1.3}_{-1.0}$         & $-0.74^{+0.35}_{-0.49}$     \\
{\boldmath$H_0$}                  & $67.20\pm 0.57$                      & $66.85\pm 0.60$             \\
{\boldmath$S_8$}                     & $0.8089\pm 0.0077$              & $0.8291\pm 0.0096$          \\
\hline
{\boldmath$\chi^2_{\rm min}$}      & $2471.861$ & $2461.405$ \\
\hline
{\boldmath $\Delta \chi^2 $ } &  $2.553$ & $1.841$ \\

\hline
\end{tabular}
}
\end{minipage}%
\caption{\small Constraints at 68\% CL on the cosmological model parameters for different data combinations. Relative $\Delta\chi^2 \equiv \chi^2_{\rm model+nCC}-\chi^2_{\rm model}$ for $w$CDM and CPLCDM models with and without nCC contribution.}
\label{tab:result}
\end{table}

\begin{figure}[t]
    \centering
    \includegraphics[width=0.8\linewidth]{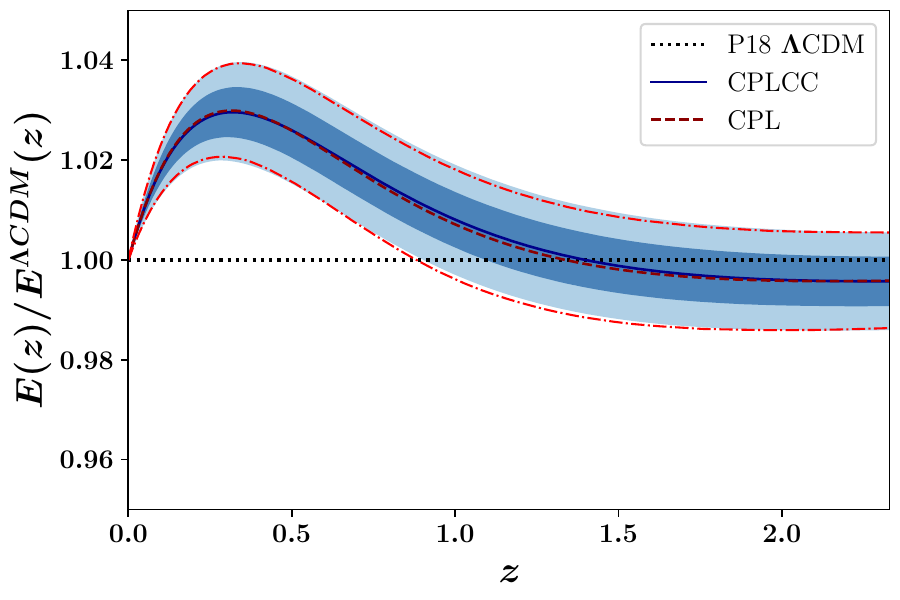}
\caption{Evolution of $E(z)/E^{\Lambda \rm CDM}(z)$ showing a comparison between CPLCCCDM vs CPLCDM at $1\sigma-2\sigma$ CL with P-ACT+DESI+DESY5 data.\vspace{-0.2cm}}
    \label{fig:Ez_cplcc_vs_cpl}
\end{figure}

\begin{figure*}[htb]
    \centering
    \begin{subfigure}[b]{0.325\linewidth}
        \includegraphics[width=\linewidth]{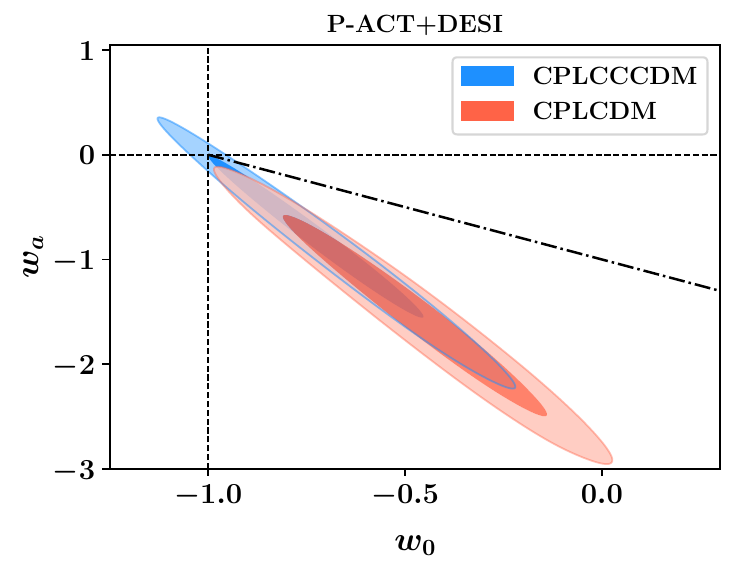}
     \caption{}
        \label{fig:w0wa1}
    \end{subfigure}
    \hfill
    \begin{subfigure}[b]{0.325\linewidth}
        \includegraphics[width=\linewidth]{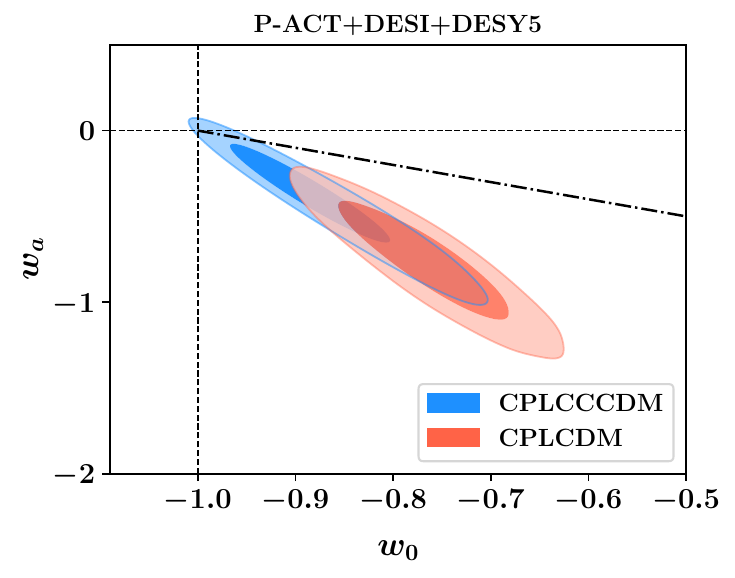}
     \caption{}
        \label{fig:w0wa2}
    \end{subfigure}
    \hfill
    \begin{subfigure}[b]{0.325\linewidth}
        \includegraphics[width=\linewidth]{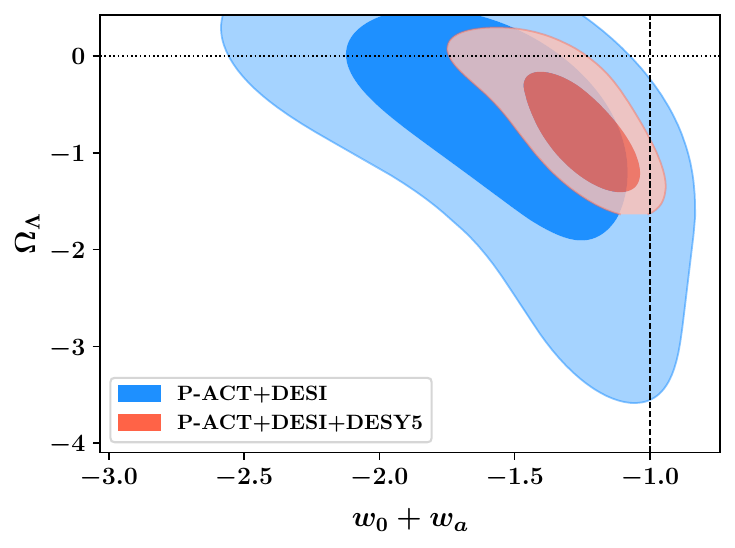}
     \caption{}
        \label{fig:w0wa4}
    \end{subfigure}
     \caption{\small 2D confidence contours at 68\% and 95\% CL for $w_0$ and $w_a$, to compare between CPLCDM and CPLCCCDM with: (a) {P-ACT}+DESI, (b) {P-ACT}+DESI+{DESY5} data. (c) Plot for $w_0 + w_a$ vs. $\Omega_\Lambda$ for CPLCCDM. $\Omega_\Lambda=0$ denotes CPLCDM. \vspace{-0.1cm}}
    \label{fig:w0wa}
\end{figure*}

\section{Data and Methodology} \label{sec:data}

\noindent We use the following datasets in our analysis:
\begin{itemize}[left=0pt]
\item[$-$] \textbf{P-ACT}: Combination of ACT DR6 \cite{ACT:2025fju} with a $Planck_{\rm cut}$ likelihood built from Planck PR3 \cite{Planck:2018vyg}, using TT spectra for $\ell < 1000$, TE/EE for $\ell < 600$, the low-$\ell$ temperature likelihood, and Sroll2 for low-$\ell$ polarization. 
\item[$-$] \textbf{DESI}: 13 DESI-BAO DR2 \cite{DESI:2025zgx} measurements across the redshift range $0.1 < z < 4.2$, including volume-averaged distance $D_V(z)/r_d$, angular diameter distance $D_M(z)/r_d$, and comoving Hubble distance $D_H(z)/r_d$, where $r_d$ is the sound horizon at the drag epoch. 
\item[$-$] \textbf{DESY5}: The DES-SN5YR sample \cite{DES:2024jxu,DES:2024upw,the_des_sn_working_group_2024_12720778}, comprising 1635 photometrically classified SN-Ia in the range $0.1 < z < 1.13$, complemented by 194 low-$z$ SN-Ia in the range $0.025 < z < 0.1$.
\item[$-$] \textbf{KiDS}: Weak lensing (WL) data from the KiDS-1000 \cite{KiDS:2020suj} survey, analyzed using COSEBIs \cite{Schneider_2010} to separate E-mode and B-mode contributions. 
\end{itemize}
The numerical analysis is conducted using the Boltzmann solver \texttt{CLASS} \cite{Blas:2011rf}, with parameter constraints derived via MCMC from \texttt{MontePython} \cite{Audren:2012wb, Brinckmann:2018cvx} and \texttt{Cobaya} \cite{Torrado:2020dgo}, assuming uniform flat priors on the model parameters 
$\left\lbrace \Omega_{\rm b} h^2, \,  \Omega_{\rm c} h^2, \, \tau, \, n_s , \, \log[10^{10}A_{s}] ,  \, 100 \, \theta_{\rm \star} , \,  w_0, \,  w_a , \, \Omega_\phi \right\rbrace$. 
A Gelman-Rubin convergence criterion of $R - 1 < 0.03$ ensures the convergence of the chains. Results are visualized using \texttt{GetDist} \cite{Lewis:2019xzd}, and Bayesian evidence is computed using \texttt{MultiNest} \cite{Feroz:2008xx, Feroz:2013hea}. \smallskip

\begin{figure*}[t]
    \centering
    \begin{subfigure}[b]{0.37\linewidth}
    \includegraphics[width=\linewidth]{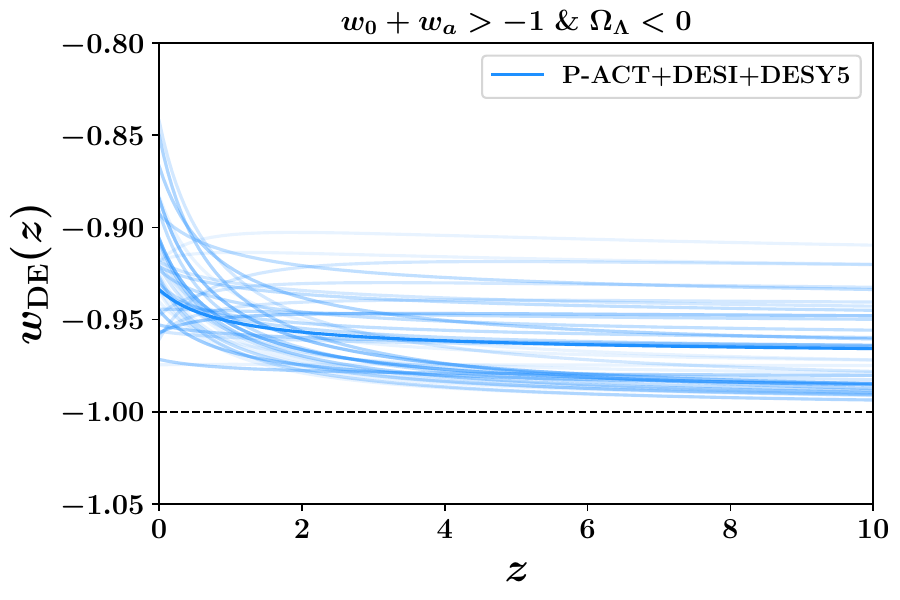}
     \caption{}
        \label{fig:wfld}
    \end{subfigure}
    % \hfill
    \begin{subfigure}[b]{0.37\linewidth}
    \includegraphics[width=\linewidth]{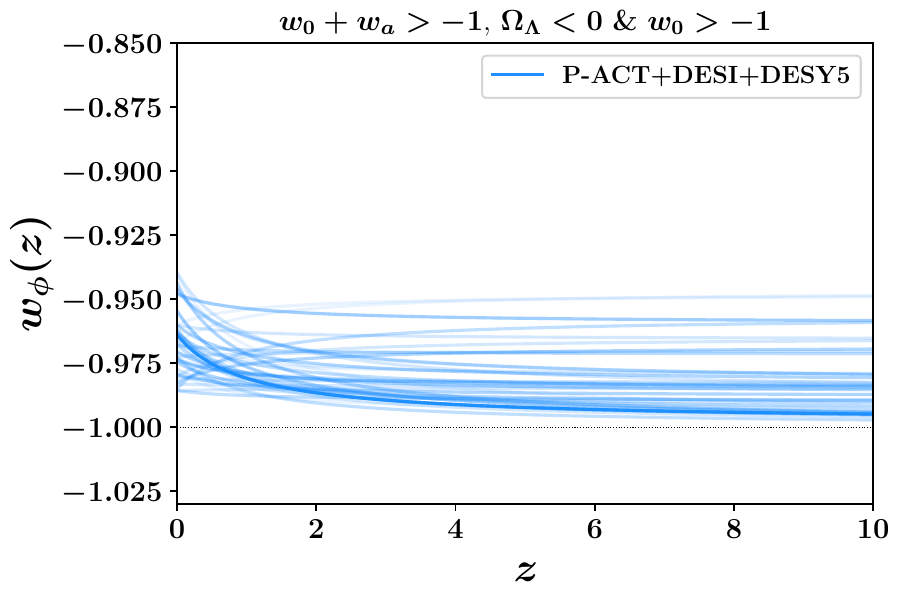}
     \caption{}
        \label{fig:w0wa2}
    \end{subfigure}
\caption{Evolution of the (a) total DE EoS $w_{\rm DE}(z)$, along with (b) the fluid's EoS, $w_\phi(z)$ for the CPLCCCDM model, derived from the {P-ACT+DESI+DESY5} chains which satisfy the non-phantom behaviour at 2$\sigma$ CL.\vspace{-0.2cm}}
    \label{fig:wz_vs_z}
\end{figure*}

\begin{figure}[t]
    \centering
    \includegraphics[width=0.7\linewidth]{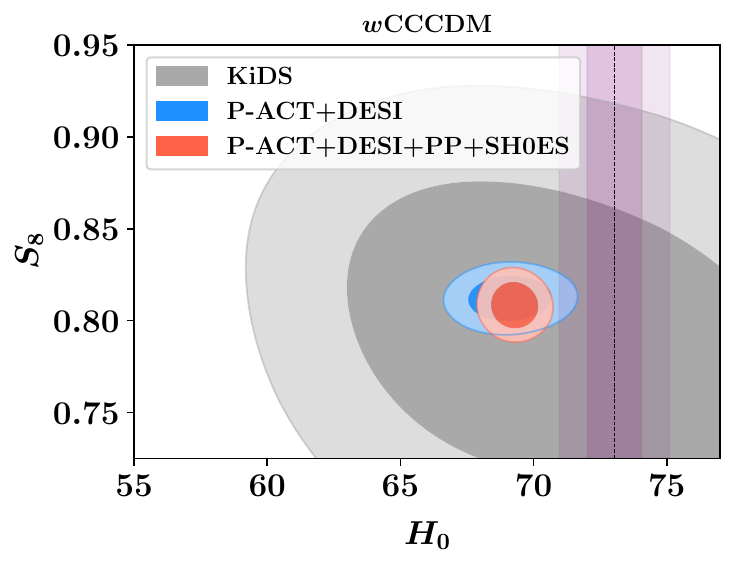}
    \caption{\small 2D confidence contours at 68\% and 95\% CL for $H_0$ and $S_8$, comparing $w$CDM and $w$CCCDM models, obtained using KiDS, and {P-ACT}+DESI data. \vspace{-0.1cm}}
    \label{fig:H0S8}
\end{figure}

\section{Results and Discussions} \label{sec:results}

We summarize the constraints on cosmological parameters at 68\% CL in Tab. \ref{tab:result}. The most important findings are as follows:

\begin{enumerate}[left=0pt]
\item The resulting constraints on the parameters, given in Table \ref{tab:result}, indicate a preference for non-zero negative values of $\Omega_\Lambda$ in the case of $w$CCCDM and CPLCCCDM models, irrespective of the data set combination. With the P-ACT+DESI+DESY5 data combination, the $w$CCCDM model shows preference for a non-zero $\Omega_\Lambda = -1.4^{+1.3}_{-1.0}$ at $1.08\sigma$ CL, whereas the  CPLCCCDM model shows preference for a non-zero $\Omega_\Lambda = -0.74^{+0.35}_{-0.49}$ at $2.12\sigma$ CL, respectively. From KiDS WL data alone, the $w$CCCDM model yeild $\Omega_\Lambda = -1.50^{+0.63}_{-0.38}$, excluding zero at $>2\sigma$ CL. This implies that the conventional $w$CDM and CPLCDM models, with $\Omega_\Lambda = 0$, are disfavored at the $1\sigma$ and $2\sigma$ CL, respectively.

\item While only the nCC results are presented here, internal checks confirm that the best-fit constraints are broadly consistent with their non-nCC counterparts (see Table \ref{tab:no_ncc_combined_results} in Appendix, also \cite{DESI:2025zgx, ACT:2025tim}), apart from shifts in parameters, $w_0$ and $w_a$, driven by the inclusion of a non-zero $\Omega_\Lambda$. 

As a representative example, in Fig.~\ref{fig:Ez_cplcc_vs_cpl} we plot the evolution of the reduced Hubble parameter $E(z)\equiv H(z)/H_0$, normalized to its $\Lambda$CDM counterpart $E^{\Lambda{\rm CDM}}(z)$, comparing the CPLCCCDM and CPLCDM models, with the shaded regions representing the $1\sigma$ and $2\sigma$ CL obtained using \textsc{P-ACT}+DESI+DESY5 data. We find that the expansion history for CPLCCCDM closely mimics CPLCDM in the range $0<z<2.33$, consistent at $2\sigma$ CL with \textsc{P-ACT}+DESI+DESY5 data. Thus, despite shifts in the DE EoS parameters, the reconstructed $E(z)$ remains nearly identical, highlighting a degeneracy in $w_0-w_a-\Omega_\Lambda$ parameter space that leaves the background dynamics indistinguishable.

%%%%%%%%%%%%%%%%%%%%%%%%%%%%%%%%%%%%%%%%%%%

\item Fig.~\ref{fig:w0wa}(a)-(b) illustrate one of the most interesting findings of this study. We plot the $w_0-w_a$ parameter space for CPLCCCDM vs CPLCDM models. The inclusion of nCC in the DE sector extends the allowed $w_0-w_a$ parameter space in the non-phantom region of the $w_0+w_a=-1$ line, which, although small, is finite. For better visualization, we plot the contour for $w_0 + w_a$ vs. $\Omega_\Lambda$ in Fig. \ref{fig:w0wa4}. Here, the $\Omega_\Lambda = 0$ line corresponds to the CPLCDM model, while the $w_0 + w_a = -1$ line marks the phantom vs non-phantom divide within the CPL parametrization. Here, note that the CPL parametrization provides a phenomenological description of the dark-energy equation of state over the redshift range probed by the data, and that the interpretation of regions in the $(w_0,\,w_a)$ plane should be treated with appropriate caution when relating them to the underlying microphysical behavior of dark energy \cite{Shlivko:2024llw, Wolf:2024eph}. Within this framework, the CPLCCCDM model allows for a small quintessence region at $2\sigma$, when the contribution from nCC to the DE sector is included, regardless of the data set under consideration.

\item In Fig. \ref{fig:wz_vs_z}, we present the evolution of the total DE EoS $w_{\rm DE}(z)$, along with the fluid's EoS, $w_\phi(z) \equiv w(z) = w_0 + w_a \left( \frac{z}{1+z} \right)$, derived directly from the P-ACT+DESI+DESY5 chains for CPLCCCDM which lies in the non-phantom region. The plot illustrates that the evolving part of the DE sector, which is contributed by the CPL/fluid parameterization of CPLCCCDM, strictly remains non-phantom under the conditions: $w_0 + w_a > -1$, $\Omega_\Lambda < 0$, and $w_0 > -1$ for all time. However, the total dark energy evolution can accommodate both phantom and non-phantom behaviour.

\item The $w_0$–$w_a$ constraints from the CPLCDM model indicate parameter regions that, when extrapolated within the CPL parametrization, correspond to an effective transition from early-phantom-like to late non-phantom behaviour with a phantom-to-non-phantom crossing at later times. Such early-time phantom-like regions are known to be theoretically problematic, leading to ghost instabilities, due to potential violations of the null energy condition (NEC) \cite{Arefeva:2006ido, Vafa:2005ui}. The presence of an nCC contribution reshapes the $w_0$–$w_a$ parameter space, opening a small yet finite region consistent with non-phantom evolution that exhibits quintessence behaviour, which can be theoretically realised using scalar field models based on field theory, satisfying the NEC \cite{Ratra:1987rm, Caldwell:1997ii}. Note that, as shown in Fig.~\ref{fig:Ez_cplcc_vs_cpl}, these parameter shifts in CPLCCDM with nCC exhibit essentially the same reconstructed $E(z)$ behaviour as in the corresponding CPLCDM case.

%%%%%%%%%%%%%%%%%%%%%%%%%%%%%%%%%%%%%%%%%%%

\item The constraints on $H_0$ for both nCC models, are similar to their respective non-nCC counterparts (see \cite{DESI:2025zgx}). Adding an extra parameter, $\Omega_\Lambda$, to account for the nCC in the DE sector results in slightly relaxed confidence intervals compared to models where $\Omega_\Lambda = 0$. Importantly, there is no deterioration in the constraints for the nCC models as far as the mean values are concerned.

\item The \textsc{P-ACT}+DESI analysis indicates that the CPLCCCDM model fails to significantly address the Hubble tension (see Tab. \ref{tab:result}). With the addition of DESY5, the mean $H_0$ shifts even lower---consistent with earlier findings for CPLCDM \cite{DESI:2025zgx}---showing that dynamical CPL parameterizations are generally less effective in alleviating the $H_0$ tension.

%%%%%%%%%%%%%%%%%%%%%%%%%%%%%%%%%%%%%%%%%%%

\begin{figure}[t]
    \centering
    \includegraphics[width=0.75\linewidth]{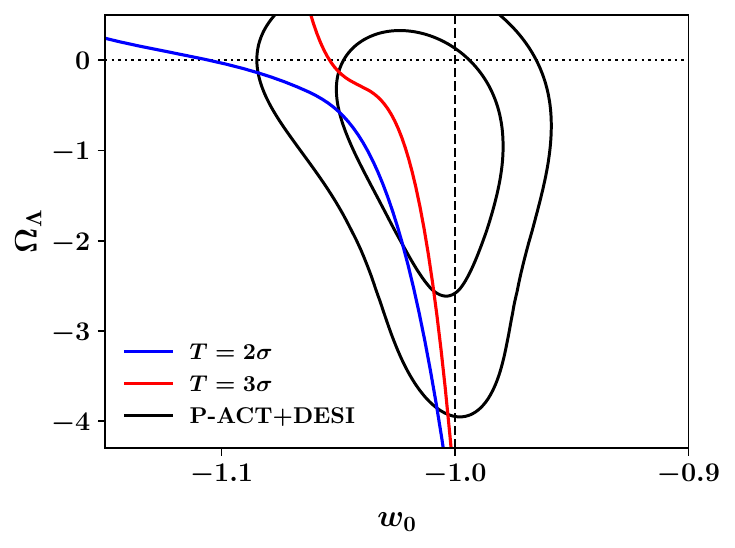}
    \caption{The $w_0$--$\Omega_{\Lambda}$ parameter space for $w$CCCDM with P-ACT+DESI where $\Omega_{\Lambda}=0$ recovers $w$CDM.}
    \label{fig:Tsw0OL_wcc}
\end{figure}

\item However, the \textsc{P-ACT}+DESI analysis shows that the $w$CCCDM model better addresses the Hubble tension, yielding $H_0$ constraints in improved agreement with the SH0ES value $H_0 = 73.01 \pm 1.04$ km Mpc$^{-1}$ s$^{-1}$ \cite{Riess:2021jrx}. In the $H_0$–$S_8$ plane (Fig.~\ref{fig:H0S8}) achieves $1\sigma$ overlap across both KiDS and \textsc{P-ACT}+DESI data, thereby simultaneously easing the $H_0$ and $S_8$ tensions.

\item Fig.~\ref{fig:Tsw0OL_wcc} shows the $w_0$--$\Omega_\Lambda$ parameter space for the $w$CCCDM model from the \textsc{P-ACT}+DESI analysis. Allowing $\Omega_\Lambda < 0$ opens regions where the $H_0$ tension with SH0ES drops below $3\sigma$, while remaining close to $w_0 = -1$. Notably, a finite portion of this parameter space even lies within the $1\sigma$ confidence contours, where the $H_0$ tension is reduced further below $2\sigma$ CL. 

\item In contrast, the restricted subspace of the $w$CCCDM chains where $\Omega_\Lambda = 0$ (the $w$CDM equivalent scenario) allows no such regions: within the $1\sigma$ contours the tension with SH0ES exceeds $3\sigma$, leaving the parameter space in significant disagreement. This underlines that permitting negative $\Omega_\Lambda$ in the full $w$CCCDM framework opens viable regions absent in its $\Omega_\Lambda=0$ limit, demonstrating that $w$CCCDM provides a more robust parameter space for reconciling the $H_0$ tension.

\item For $S_8$, the KiDS analysis indicates that nCC models favor values consistent with Planck CMB measurements, as well as with \textsc{P-ACT}+DESI, and \textsc{P-ACT}+DESI+DESY5. This demonstrates that the nCC framework enhances structure formation, accommodating $S_8$ values from both WL and CMB-anchored constraints in better agreement.

%%%%%%%%%%%%%%%%%%%%%%%%%%%%%%%%%%%%%%%%%%%

\begin{figure}[t]
    \centering
    \begin{subfigure}[b]{0.495\linewidth}
    \includegraphics[width=\linewidth]{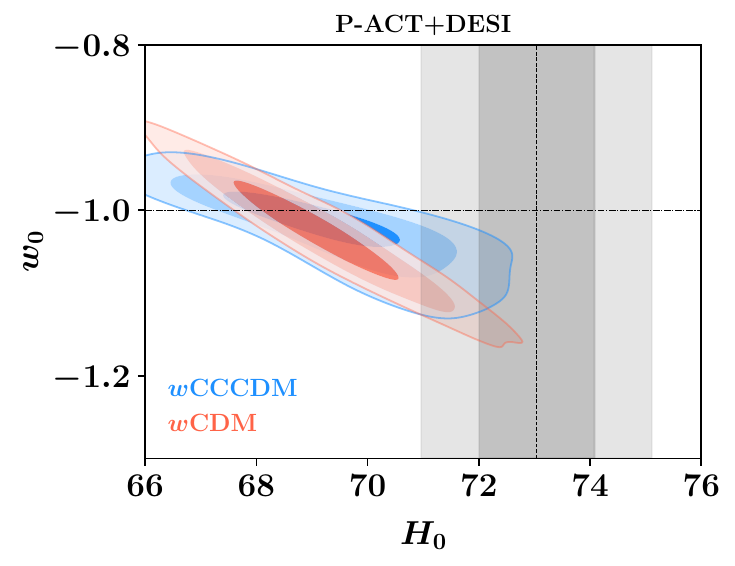}
    \caption{}
    \label{fig:H0w0_wcc}
    \end{subfigure}
    \begin{subfigure}[b]{0.495\linewidth}
    \includegraphics[width=\linewidth]{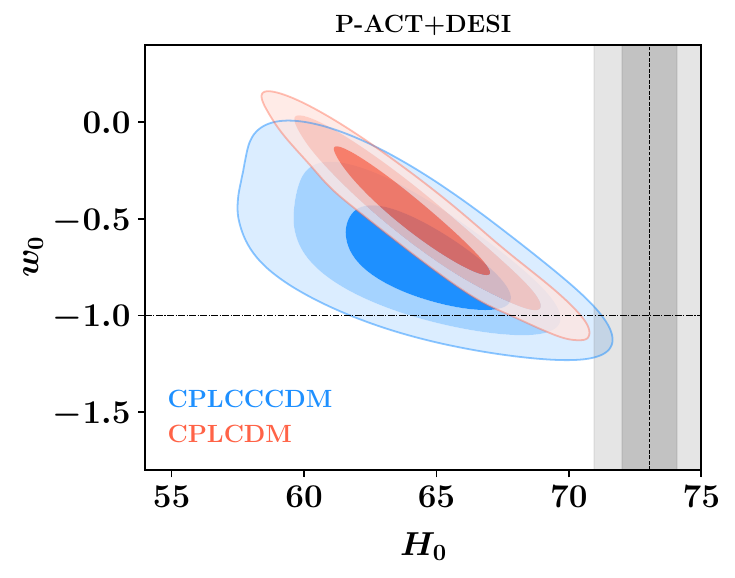}
    \caption{}
    \label{fig:H0w0_cplcc}
    \end{subfigure}
    \caption{\small 2D confidence contours at 68\% and 95\% CL for $H_0$ and $w_0$, illustrating the impact of including nCC by comparing: (a) $w$CCCDM vs $w$CDM, (b) CPLCCCDM vs CPLCDM, for P-ACT+DESI data. \vspace{-0.1cm}}
    \label{fig:H0w0}
\end{figure}

\item In the $w$CCCDM model (Fig. \ref{fig:H0w0_wcc}), a significant portion of the $w_0–H_0$ parameter space allows quintessence-like behavior ($w_0 > -1$) within 1$\sigma$ CL, than compared to $w$CDM . However, addressing the Hubble tension in this framework requires moving into the phantom regime ($w_0 < -1$), which is theoretically less appealing. However, for the CPLCCCDM model (Fig. \ref{fig:H0w0_cplcc}), the majority of the $w_0–H_0$ parameter space lies in the non-phantom region for the \textsc{P-ACT}+DESI combination. Thus, dynamical fluid parametrizations are physically more viable compared to the constant fluid parameterizations.

\item To quantify the relative difference between models with and without an nCC, we report in Table~\ref{tab:result} the relative change in $\chi^2_{\rm min}$ values, defined as $\Delta\chi^2 \equiv \chi^2_{\rm model+nCC}-\chi^2_{\rm model}$, where negative (positive) values indicate a marginally better (worse) fit upon inclusion of nCC. The resulting $\Delta\chi^2$ values $\approx \pm 2$ units and vary in sign depending on the dataset, implying that both scenarios provide comparably good descriptions of the data. Given the additional degree of freedom introduced by nCC and the large number of data points, the associated fractional change in $\Delta\chi^2$ per degree of freedom is negligible. We therefore conclude that current observations do not statistically distinguish between the nCC and non-nCC models, and that the inclusion of nCC does not degrade the quality of the fit.

\end{enumerate}

\begin{figure*}[t]
    \centering
    \begin{subfigure}[b]{0.3\linewidth}
        \includegraphics[width=\linewidth]{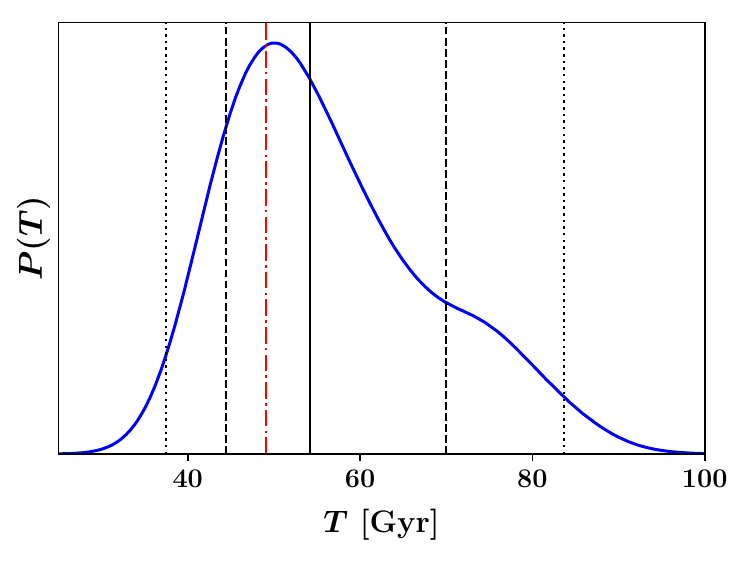}
     \caption{}
        \label{fig:lifespan_wcc}
    \end{subfigure}
    \hfill
    \begin{subfigure}[b]{0.3\linewidth}
        \includegraphics[width=\linewidth]{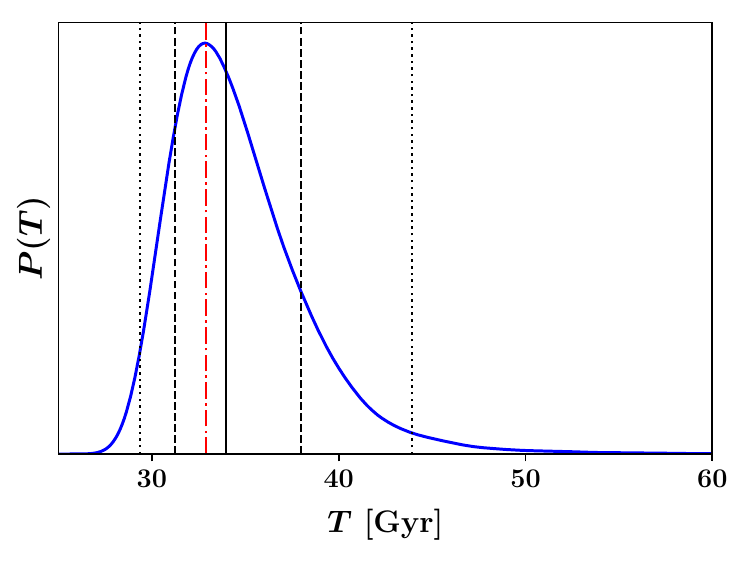}
     \caption{}
        \label{fig:lifespan_cplcc}
    \end{subfigure}
    \hfill
    \begin{subfigure}[b]{0.335\linewidth}
        \includegraphics[width=\linewidth]{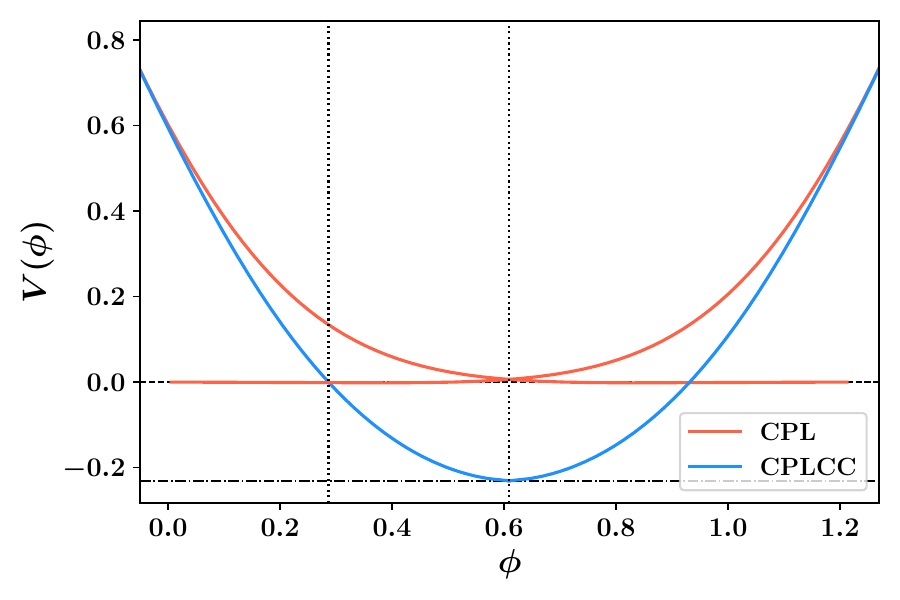}
     \caption{}
        \label{fig:Vphi_vs_phi}
    \end{subfigure}
\caption{{Posterior distribution of the total lifespan of the universe $\boldsymbol{T}$ (in Gyrs), assuming (a) $w$CCCDM and (b) CPLCCCDM using \textsc{P-ACT}+\textsc{DESI}+\textsc{DESY5} chains. (c) Reconstruction of $V(\phi)$ vs $\phi$ for CPLCC vs CPL. } \vspace{-0.2cm}}
    \label{fig:Vphi_lifespan_cplcc}
\end{figure*}

\noindent \textit{\underline{Lifespan of the Universe}}: On extending the $w$CCCDM and CPLCCCDM models into the \emph{future regime} ($z < 0$), one can identify a critical redshift $z_c$, defined by $E(z_c)=0$, where expansion halts, and re-collapse begins due to the effective nCC term. The total cosmic lifetime is then given by,  
    $\boldsymbol{T} = t_0 + \Delta t_{\rm c} \, ,$ where   \begin{equation}
     t_0 = \int_0^\infty \frac{{\rm d} z}{(1+z)H(z)} \, ;  ~~\Delta t_{\rm c} = \int_{z_c}^{0} \frac{{\rm d}z}{(1+z)H(z)} \, .   
    \end{equation}
Using \textsc{P-ACT}+DESI+DESY5, the $w$CCCDM and CPLCCCDM models yield a finite lifetime of $\boldsymbol{T} = 54.175^{+15.828}_{-9.747}$ Gyr and $\boldsymbol{T} = 33.985^{+3.980}_{-2.726}$ Gyr (shown in Fig. \ref{fig:lifespan_wcc}-\ref{fig:lifespan_cplcc}), respectively. Restricting the $w$CCCDM calculation with \textsc{P-ACT}+DESI to samples with $<2\sigma$ tension with SH0ES approximates this estimate to $\boldsymbol{T} \sim 55.159$ Gyr. Thus, both nCC frameworks predict that the Universe has a finite lifespan. \\

\noindent \textit{\underline{Reconstruction of Effective Potential}}: {In Fig. \ref{fig:Vphi_vs_phi}, we reconstruct the potential $V(\phi)$ as a function of the scalar field $\phi$ from its kinetic and potential contributions, capturing the dynamics of both CPL and CPLCC DE parametrizations:
\begin{equation}
\begin{split}
\dot{\phi}^2(z) &= \tfrac{1}{2}\,[1 + w_{\rm DE}(z)]\,\rho_{\rm DE}(z)\, , \\
V(z) &= \tfrac{1}{2}\,[1 - w_{\rm DE}(z)]\,\rho_{\rm DE}(z)\, .
\end{split}
\end{equation}
By numerically solving $\frac{d\phi}{da} = \frac{\dot{\phi}}{aH(a)}$ with present-epoch initial conditions ($a(t_0) = a_0 \equiv 1$, $\phi(a_0) = \phi_0 \equiv 0$), we can map $\phi(a)$ to the reconstructed $V(a)$. The presence of nCC shifts the potential to a negative minimum in the CPLCC case, driving late-time re-collapse with a finite cosmic lifetime, whereas in the CPL case $\phi$ slowly rolls to $0$, allowing for indefinite expansion. 
}

\section{Concluding Remarks} \label{sec:conclusion}

\noindent Thus, \textit{our analysis shows that the presence of an nCC in the DE sector is compatible with current observational constraints, incorporating an additional nCC component in the DE sector offers a significant advantage in modelling, as it enables a region in the $w_0$–$w_a$ parameter space that aligns with a non-phantom DE EoS. Moreover, a non-phantom behaviour can be physically well-motivated and is theoretically consistent, thereby facilitating more robust and reliable modelling of the DE sector. Within the constant-$w$ framework, presence of nCC helps in alleviating the Hubble tension. Allowing for an nCC component further implies the possibility of a finite cosmic lifetime, with tight bounds from observations, leading to an eventual re-collapse, with profound implications for the fate of the universe.} \\

Finally, in light of the phenomenological viability of nCC scenarios in the cosmic dark sector, upcoming observational galaxy surveys such as Euclid \cite{Amendola:2016saw}, the Vera Rubin Observatory \cite{Blum:2022dxi}, and next-generation CMB \cite{LiteBIRD:2020khw, CMB-S4:2016ple} and 21 cm experiments \cite{Dutta:2022pwh}, will play a critical role in testing the predictions of nCC models \cite{Dash:2023scq}. These surveys are set to provide more precise constraints on the DE sector, presenting an opportunity to confirm or refute evidence for the presence of an additional nCC component in the DE sector in shaping cosmic expansion and structure formation. 

The connection between AdS vacuum and string theory further strengthens the theoretical appeal of this framework. String theory naturally accommodates nCC through its AdS vacuum structure, supported by the AdS/CFT correspondence \cite{Demirtas:2021ote, VanRaamsdonk:2023ion}. This linkage underscores the potential of the nCC models to bridge cosmological observations with high-energy physics, paving the way for a more unified understanding of the universe.  \\

\begin{acknowledgments}
PM acknowledges funding from the Anusandhan National Research Foundation (ANRF), Govt of India under the National Post-Doctoral Fellowship (File no. PDF/2023/001986). DK acknowledges the Ministry of Education (MoE) fellowship at MNIT Jaipur. AAS acknowledges the funding from ANRF, Govt of India under the research grant no. CRG/2023/003984. We acknowledge the use of the HPC facility, Pegasus, at IUCAA, Pune, India.
\end{acknowledgments}

\appendix

\squeezetable
\begin{table}[t]
\setlength{\tabcolsep}{15pt}
\centering
\begin{minipage}{0.5\textwidth}
\centering
\resizebox{\textwidth}{!}{\renewcommand{\arraystretch}{1.3}
\begin{tabular}{lcc}
\hline\hline

\multicolumn{3}{c}{\textbf{KiDS}} \\
\hline
\textbf{Parameter} & \textbf{$w$CDM} & \textbf{CPLCDM} \\
\hline
{\boldmath$\omega_{\rm cdm}$}        & $0.123\pm 0.043$            & $0.119^{+0.029}_{-0.049}$   \\
{\boldmath$\omega_{\rm b}$}         & $0.0225\pm 0.0020$          & $0.0225\pm 0.0020$          \\
{\boldmath$h$}                      & $0.728\pm 0.047$            & $0.735\pm 0.052$            \\
{\boldmath$\sigma_{8,0}$}           & $0.81\pm 0.13$              & $0.83\pm 0.12$              \\
{\boldmath$n_s$}                    & $0.958\pm 0.069$            & $0.934^{+0.032}_{-0.090}$   \\
{\boldmath$w_0$}                    & $-0.98^{+0.36}_{-0.22}$     & $-0.93^{+0.59}_{-0.17}$     \\
{\boldmath$w_a$}                    &      $-$                       & $-0.60^{+0.93}_{-0.62}$     \\
\hline
{\boldmath$\Omega_{\rm m}$}         & $0.278^{+0.070}_{-0.099}$   & $0.263^{+0.060}_{-0.084}$   \\
{\boldmath$H_0$}                      & $72.8\pm 4.7$            & $73.5\pm 5.2$            \\
{\boldmath$S_8$}                    & $0.756^{+0.038}_{-0.042}$   & $0.760\pm 0.053$            \\
\hline
{\boldmath$\chi^2_{\rm min}$}      & $82.936$ & $83.250$ \\
\hline\hline

\multicolumn{3}{c}{\textbf{P-ACT+DESI}} \\
\hline
\textbf{Parameter} & \textbf{$w$CDM} & \textbf{CPLCDM} \\
\hline
{\boldmath$\omega_{\rm b}$}         & $0.02254\pm 0.00011$        & $0.02249\pm 0.00011$        \\
{\boldmath$\omega_{\rm cdm}$}       & $0.11760\pm 0.00094$        & $0.1190\pm 0.0011$          \\
{\boldmath$100\,\theta_s$}          & $1.04179\pm 0.00025$        & $1.04169\pm 0.00025$        \\
{\boldmath$\ln(10^{10}A_s)$}         & $3.061\pm 0.011$            & $3.052^{+0.011}_{-0.012}$   \\
{\boldmath$n_s$}                    & $0.9745\pm 0.0035$          & $0.9712\pm 0.0035$          \\
{\boldmath$\tau_{\rm reio}$}        & $0.0609^{+0.0056}_{-0.0070}$& $0.0594^{+0.0057}_{-0.0068}$\\
{\boldmath$w_0$}                    & $-1.025\pm 0.039$           & $-0.47\pm 0.21$             \\
{\boldmath$w_a$}                    &         $-$                    & $-1.54\pm 0.61$             \\
\hline
{\boldmath$\Omega_{\rm m}$}         & $0.2949\pm 0.0077$          & $0.347\pm 0.022$            \\
{\boldmath$\Omega_{\rm \phi}$}      & $0.7050\pm 0.0077$          & $0.653\pm  0.022$           \\
{\boldmath$H_0$}                    & $69.11\pm 0.99$             & $64.1^{+1.8}_{-2.1}$        \\
% {\boldmath$\sigma_{8,0}$}           & $0.819\pm 0.013$            & $0.785\pm 0.017$            \\
{\boldmath$S_8$}                    & $0.8118\pm 0.0078$          & $0.843\pm 0.014$            \\
\hline
{\boldmath$\chi^2_{\rm min}$}      & $823.610$ & $820.281$ \\
\hline\hline

\multicolumn{3}{c}{\textbf{P-ACT+DESI+DESY5}} \\
\hline
\textbf{Parameter} & \textbf{$w$CDM} & \textbf{CPLCDM} \\
\hline
{\boldmath$\omega_{\rm b}$}         & $0.02258\pm 0.00011$        & $0.02250\pm 0.00011$        \\
{\boldmath$\omega_{\rm cdm}$}       & $0.11659\pm 0.00088$        & $0.11865\pm 0.00095$        \\
{\boldmath$100\,\theta_s$}          & $1.04185\pm 0.00025$        & $1.04172\pm 0.00025$        \\
{\boldmath$\ln(10^{10}A_s)$}         & $3.068\pm 0.012$            & $3.054\pm 0.011$            \\
{\boldmath$n_s$}                    & $0.9767\pm 0.0034$          & $0.9721\pm 0.0033$          \\
{\boldmath$\tau_{\rm reio}$}        & $0.0624^{+0.0059}_{-0.0073}$& $0.0596^{+0.0057}_{-0.0064}$\\
{\boldmath$w_0$}                    & $-0.954\pm 0.023$           & $-0.765\pm 0.058$           \\
{\boldmath$w_a$}                    &         $-$                    & $-0.77\pm 0.23$             \\
\hline
{\boldmath$\Omega_{\rm m}$}         & $0.3089\pm 0.0052$          & $0.3173\pm 0.0058$          \\
{\boldmath$\Omega_{\rm \phi}$}      & $0.6910\pm 0.0052$          & $0.6826\pm 0.0058$          \\
{\boldmath$H_0$}                    & $67.28\pm 0.57$             & $66.85\pm 0.59$             \\
% {\boldmath$\sigma_{8,0}$}           & $0.7976\pm 0.0087$          & $0.8064\pm 0.0084$          \\
{\boldmath$S_8$}                    & $0.8093\pm 0.0082$          & $0.8293\pm  0.0091$         \\
\hline
{\boldmath$\chi^2_{\rm min}$}      & $2469.308$ & $2459.564$ \\
\hline\hline

\end{tabular}
}
\end{minipage}%
\caption{\small Constraints at 68\% CL on parameters for KiDS, P-ACT+DESI, and P-ACT+DESI+DESY5, comparing $w$CDM and CPLCDM.}
\label{tab:no_ncc_combined_results}
\end{table}

%\nocite{*}

\bibliography{references}

\end{document}